\documentclass[onecolumn,journal]{IEEEtran}

\usepackage{epsfig}
\usepackage{amsmath}
\usepackage{theorem}
\usepackage{amsmath}
\usepackage{mathrsfs}
\usepackage{amsfonts}
\usepackage{amssymb}
\usepackage{mathrsfs}
\usepackage{euscript}
\usepackage{graphicx}
\usepackage{multirow}
\usepackage{cite}
\usepackage{url}
\usepackage{algorithm}

\renewcommand{\baselinestretch}{1.5}

\begin{document}

\title{ Power Control and Scheduling In Low SNR Region In The Uplink of Two Cell Networks  }

\author{
\IEEEauthorblockN {Ataollah khalili, Soroush Akhlaghi\\}
\IEEEauthorblockA {Department of Engineering,
Shahed University, Tehran, Iran\\
Emails: \{a.khalili,~akhlaghi\}@shahed.ac.ir
}
}
% make the title area

\maketitle

\begin{abstract}
In this paper we investigate the sub-channel assignment and power control to maximize the total sum rate in the uplink of two-cell network. It is assumed that there are some sub-channels in each cell which should be allocated among some users. Also, each user is subjected to a power constraint.

    The underlying problem is a non-convex mixed integer non-linear optimization problem which does not have a trivial solution. To solve the problem, having fixed the consumed power of each user, and assuming low co-channel interference region, the sub-channel allocation problem is reformulate into a more mathematically tractable problem which is shown can be tackled through the so-called Hungarian algorithm.
Then, the consumed power of each user is reformulated as a quadratic fractional problem which can be numerically derived. Numerical results demonstrate the superiority of the proposed method in low SNR region as compared to existing works addressed in the literature
\end{abstract}
\section{Introduction}\label{sec:intro}
Resource allocation is considered as a major challenge in wireless communication networks. This is due to the limited available bandwidth, and total power, while there is a non-stop demand for emerging communication services. It is widely recognized that Orthogonal Frequency Division Multiple Access (OFDMA) can effectively divide the available bandwidth into orthogonal sub-channels to be allocated among active users. OFDMA, on the other hand, can be employed in multi-user multi-path dispersive channel as it can effectively divide a frequency selective fading channel into some narrowband flat fading channels \cite{b1}. Noting this, OFDMA technique has been widely adopted in broadband wireless communications over the last decade, due to its flexibility in resource allocation.

It is worth mentioning that in an OFDMA system, the intra-cell interference is simply avoided due to the orthogonality among sub-channels \cite{b2,b3,b4,b5,b6,b7}. In conventional cellular system, the user from different place may utilize the same frequency, which will cause the inter-cell interference. the One limiting factor that influences cellular performance is the interference from neighbor cells, the so called Inter-Cell Interference (ICI). To mitigate interference effectively, a classical method is to adjust frequency reuse factor, however this technique alleviate inter-cell interference it can reduce the available spectrum within each cell and may degrade the overall throughput. Nevertheless, future mobile network, such as LTE network require to support high data rate service. So to this overcome this problem the Fractional Frequency Reuse (FFR) has been proposed as a technique, since it can efficiently employ the available frequency spectrum. In this paper we consider an uplink two cell system which share the whole available spectrum. The frequency reuse factor is one.
The problem of assigning sub-channels and allocating power to users in an OFDMA system has attracted many attentions in recent years. Moreover, the resource allocation problem in the uplink is more challenging than that of the downlink as the uplink interference is mostly affected by neighbouring co-channel users \cite{b7,b8}.

In this regard, a plethora of works are devoted to explore effective ways of assigning sub-channels as well as allocating power to optimize a performance function \cite{b3,b4,b5,b6,b7,b8,b9,b10,b11}. For instance, the author of \cite{b3} investigate the joint sub-channel assignment and power control mechanism in terms of maximizing the sum rate in an uplink OFDMA network. This problem is non-convex mixed integer non-linear problem which can be solved by adding a penalty term to the objective function and relax the integer variables can be converted into a difference of two concave function (DC) problem. It is worth mentioning that the sub-optimal problem derived in\cite{b3} is too complex and the authors do not provide the condition under which the proposed method approaches the optimal solution.

The author of \cite{b4}determines the resource allocation in multi-cell OFDMA networks in order to jointly optimize the energy efficiency and spectral efficiency performance which allocate the sub-channel and power iteratively. This method, however, suffers from poor performance as compared to \cite{b3}.

In \cite{b5} the joint sub-channel assignment and power control problems in a cellular network with the objective of enhancing the quality of-service is studied. Accordingly, this problem is tackled in two steps. First, it attempts to assign the sub-channels assuming all users make use of an equal power. Then, the power of each user is optimized for the assigned channels. Again, this problem has a poor performance as compared to \cite{b3}.

For achievement to the maximum throughput in this network. We consider joint power and sub-channel allocation in the uplink of OFDMA network. The optimization problem is a highly non-convex mixed integer non-linear problem. The suboptimal power and resource allocation policy can solve the considered problem via an optimization approach based on the sub-gradient method. Also, a low-complexity suboptimal algorithm based on the Hungarian method is proposed and it is shown that its results is close to optimal. Thus the main contribution of this paper can be summarized as follows: we are going to propose a novel approach which performs well in low SNR region. To this end, it is demonstrated that the original problem can be simplified to an assignment problem in low SNR region which can be effectively tackled through using the so-called Hungarian method \cite{b13}. More specifically, for two-cell network and every sub-channel, a pair of users each belonging to one of existing cells is identified. Then, the power of each pair of users is addressed through converting the original problem into a fractional quadratic problem, hence, the associated power can be effectively computed.
\begin{figure}
   \includegraphics[width=6.00cm]{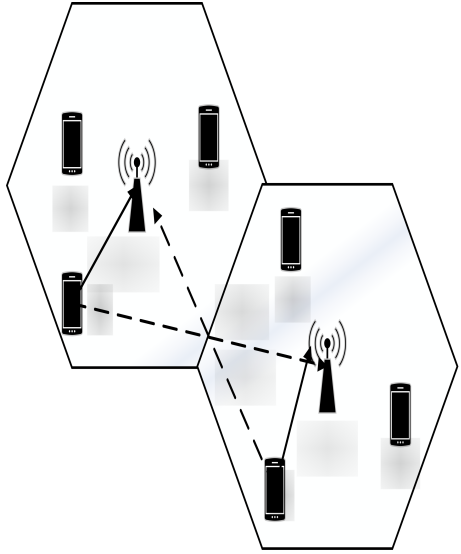}
    \centering
    \centering \caption{Structure of the considered network.}
  \label{fig:system_model}
\end{figure}
\section{System Model}\label{sec:model}
In this paper, we consider a two-cell network composing of users per cell trying to send their signals to the base station in the uplink channel. Also, there are $N$ sub-channels to be used by each user pair such that the signal arising from a co-channel user in one cell is treated as an interference in the other cell. The set of users in the $j^{th}$ cell is represented by$I_{j}$ where $j={1,2}$.
The network structure is depicted in Fig. 1, where the dotted lines represent inter-cell interference paths. Moreover, the uplink transmission from the $i^{th}$ user in the $j^{th}$ cell to the base station$j$ goes through a Rayleigh flat fading channel denoted by , $g^{n}_{i(j)j}$.where the channel strength is denoted by$h^{n}_{i(j)j}=|g^{n}_{i(j)j}|^{2}$
$p^{n}_{ij}$ and $x^{n}_{ij}$ , respectively represent the transmit power and a zero/one indicator showing if the$n^{th}$ sub-channel is assigned to$i^{th}$  user residing in cell $j$ where $\mathcal{N}=\{1,2,...N\}$  . It is worth mentioning that channel state information is globally known at the base stations.
The data rate of the user located in the cell using the sub-channel according to the Shannon capacity formula can be mathematically written as follows:

\begin{eqnarray}\label{sysmod1}
R^{n}_{ij}=\log_2(1+\frac{ p^{n}_{ij} h^{n}_{i_(j),j}}{\sigma^2+\sum_{j'\neq j}\sum_{k\in I_{j'}}  x^n_{kj'} p^{n}_{kj'} h^{n}_{k(j'),j}})
\end{eqnarray}
Where $h^{n}_{i_(j),j}$ is indicate the channel from user $i^{th}$ in the $j^{th}$ cell to the$j^{th}$  cell and $h^{n}_{k(j'),j}$ indicates the interference channel from the user in the cell on the cell . Also, $\sigma^2$ is additive white Gaussian noise power.Moreover,in the studied network, the set of users taking the same sub-channel are named as user pairs, where each user pair consists of one user in each cell, so its cardinality is equal to the number of cells. For the studied model, since the cell number is two, each two users having the same sub-channel from different cells make a user pair. Moreover, since different user pairs make use of different orthogonal sub-channels, the user pairs do not get interference from other pairs and their interference is just depend on user(s) inside the pair.
\section{Problem Formulation}
The aim of this section is to find sub-channel assignment and optimal power allocation such that the sum-rate of the network is maximized. Mathematically speaking, the optimization problem is written as follows:
 \begin{equation}\label{max_prob.2}
\begin{aligned}
\max_\textbf{{x,p}}\sum_{j=1}\sum_{i=1}&\sum_{n=1}\log_2(1+\frac{x^n_{ij}p^{n}_{ij} h^{n}_{i_(j),j}}{\sigma^2+\sum_{j'\neq j}\sum_{k \in I_{j'}} x^n_{kj'} p^{n}_{kj'} h^{n}_{k_ (j'),j}})\\
\text{subject to}~~ &{{C}_{1}}:~~~ \underset{n=1}{\overset{N}{\mathop \sum }}\,x_{ij}^{n}p_{ij}^{n}\le {{p}_{max}}\\
&{{C}_{2}}:~p_{ij}^{n}\ge 0\\
&{{C}_{3}}:~\underset{n=1}{\overset{N}{\mathop \sum }}\,x_{ij}^{n}=1\\
&{{C}_{4}}:~x_{ij}^{n}\in \left\{ 0,1 \right\}\\
\end{aligned}
\end{equation}
Where $\textbf{x}=[\textbf{x}_{11},...,\textbf{x}_{I_ {|{j}|}j}]$ and $\textbf{p}=[\textbf{p}_{11},...,\textbf{p}_{I_ {|{j}|}j}]$  .where $\textbf{x}=[x^{1}_{ij},...,x^{n}_{ij}]$ and $\textbf{p}=[p^{1}_{ij},...,p^{n}_{ij}]$  .In (\ref{max_prob.2}), C1 represents the power constraint of each user.C2 indicates transmit powers take positive values. In addition, C3 shows that a single sub-channel should be allocated to each user and C4 indicates that the indicator takes zero/one values.
The problem of (2) is non-convex in general. We formulated the problem into a more mathematically tractable form.
It is noteworthy that since $x^{n}_{ij}$  is a binary variable we can write:
\begin{eqnarray}\label{sysmod2}
x^{n}_{ij}R^{n}_{ij}=\log_2(1+\frac{x^{n}_{ij} p^{n}_{ij} h^{n}_{i_(j),j}}{\sigma^2+\sum_{j'\neq j}\sum_{k\in I_{j'}}  x^n_{kj'} p^{n}_{kj'} h^{n}_{k(j'),j}})
\end{eqnarray}
It should be noted that as  $x^{n}_{ij}$ takes zero/one values, hence, both sides of (\ref{sysmod2}) becomes zero when $x^{n}_{ij}$ is zero. Similarly, referring to the definition of  $R^{n}_{ij}$ in (\ref{sysmod1}), equation (\ref{sysmod2}) holds for the case of $x^{n}_{ij}$ . One can readily verify that the optimization problem in (\ref{max_prob.2}) involves some continuous variables$p^{n}_{ij}$ and integer variables $x^{n}_{ij}$ , hence, it is not convex in general \cite{b12}. Thus, it does not yield to a trivial solution. However, having fixed the transmit power at its maximum allowable value, and considering low SNR region, it is shown that using some approximations, the original problem can be converted to an assignment problem leading to the best sub-channel selection. Then, the optimal powers associated with selected sub-channels are numerically derived in terms of maximizing the sum-rate of the network.
\section{Resource allocation}\label{sec:multilevel}
This section tends to determine the sub-channel assignment of a each user.It is assumed the consumed power of each user is constant value and operate in low SINR region.So,the following approximation can be used in order to simplify the rate of a user,i.e.$R^{n}_{ij}$ \begin{eqnarray}\label{sysmod4}
R^{n}_{ij}=\log_2(1+\frac{ p_{max} h^{n}_{i_(j),j}}{\sigma^2+\sum_{j'\neq j}\sum_{k\in I_{j'}}  x^n_{kj'} p_{max} h^{n}_{k(j'),j}})\approx\frac{ p_{max} h^{n}_{i_(j),j}}{\sigma^2+\sum_{j'\neq j}\sum_{k\in I_{j'}} p_{max} h^{n}_{k(j'),j}}\approx\frac{ p_{max} h^{n}_{i_(j),j}}{\sigma^2}
\end{eqnarray}
In this case noting that when SINR is much smaller than one, the log function can be approximated as . Where it is assumed that the noise power is greater than that of the interference. As a result, one can use equation (\ref {sysmod4}) as an approximated achievable rate of the $i^{th}$ user in the$j^{th}$ cell for the $n^{th}$ sub-channel. As a result, the optimization problem can be reformulated as
  \begin{equation}\label{max_prob.3}
\begin{aligned}
\max_{\textbf{x}}\sum_{j=1}\sum_{i=1}&\sum_{n=1}\ \frac{x^n_{ij}p_{max} h^{n}_{i_(j),j}}{\sigma^2}\\
\text{subject to}~~ &{{C}_{1}}:~\underset{n=1}{\overset{N}{\mathop \sum }}\,x_{ij}^{n}=1\\
&{{C}_{2}}:~x_{ij}^{n}\in \left\{ 0,1 \right\}\\
\end{aligned}
\end{equation}
This problem can be tackled through the so called assignment problem. The assignment problem is one of the fundamental combinatorial optimization problems. It consists of finding a maximum weight matching (or minimum weight perfect matching) in a weighted bipartite graph and can be solved by using the Hungarian algorithm \cite{b13}. Suppose that we have N sub-channels to be assigned to N users on a one to one basis. Also, the cost of assigning sub-channels to users are known. It is desirable to find the optimal assignment minimizing the total cost. Let’s $c^{n}_{ij}$  be the cost of assigning the $n^{th}$ sub-channel to the $i^{th}$ user. We define the $n*n$ cost matrix
 to solve the assignment problems in polynomial time,the Hungarian algorithm is suggested.The algorithm models an assignment problem as an $n*i$ cost matrix, where each element indicate the cost of allocating $n^{th}$ sub-channel to the $i^{th}$ user.Let us we define the cost matrix $C_{ni}$ to be $n*n$ matrix. As a result,the cost matrix C associated with the assignment problem is constructed such that the element of $i^{th}$row and the $j^{th}$ column,i.e,$C^{n}_{ij}$ is set to:
   \begin{equation}\label{sysmod6}
 \frac{p h^{n}_{i_(j),j}}{\sigma^2}\\
 \end{equation}
The Hungarian method find the best sub-channel which maximizes:
 \begin{equation}\label{max_prob.3}
\begin{aligned}
\max_{\textbf{x}}\sum_{j=1}\sum_{i=1}&\sum_{n=1}x^{n}_{ij}c^{n}_{ij}\\
\text{subject to}~~ &{{C}_{1}}:~\underset{n=1}{\overset{N}{\mathop \sum }}\,x_{ij}^{n}=1\\
&{{C}_{2}}:~x_{ij}^{n}\in \left\{ 0,1 \right\}\\
\end{aligned}
\end{equation}
Where $x^{n}_{ij}$ is an integer variable and indicator ensuring each sub-channel is assigned to one user pair based on the constraint  in (5). It should be noted that the assignment problem can be extended to a more general case of having N sub-channels and I users through without imposing any constraint on the size of sub-channels and users. .
\section{Power Control Strategy }\label{opt_power}
In the previous section, each sub-channel is assigned to one user pair. In the studied network, each user in a cell has the same sub-channel as another user in the neighboring cell. These two users with the same sub-channel are named as a user pair. Different user pairs have different sub-channels and their transmitted signals do not interferer on each other. Therefore, each user pair has no effect on other pairs. Thus, the total network throughput maximization problem can be simplified by the throughput maximization on each individual pair. Typically, the sum rate maximization problem for the first user pair can be written as:
 \begin{equation}\label{max_prob.4}
\begin{aligned}
\max_{\textbf{p}}&~log_2(1+\frac{ p^{n}_{11} h^{n}_{1(1),1}}{\sigma^2+p^{n}_{12} h^{n}_{1(2)1}})+log_2(1+\frac{ p^{n}_{12} h^{n}_{1(2)2}}{\sigma^2+p^{n}_{11} h^{n}_{1(1),2}})\\
\text{subject to}~~ &{{C}_{2}}:~p_{11}^{n}\ge 0 ~and ~p_{12}^{n}\ge 0\\
&{{C}_{3}}:~~~ \underset{n=1}{\overset{N}{\mathop \sum }}\,x_{ij}^{n}p_{ij}^{n}\le {{p}_{max}}\\
\end{aligned}
\end{equation}
To simplify, some variables are changed using the definitions:
\begin{equation}\label{sysmod7}
  h^{n}_{1(1),1}=a, h^{n}_{1(2),1}=b,h^{n}_{1(2),2}=c,h^{n}_{1(2),1}=d
\end{equation}
Thus, the problem in (8)is changed as follow :
\begin{equation}\label{max_prob.5}
\begin{aligned}
\max_{p^{n}_{11},p^{n}_{12}}&log_2(1+\frac{ p^{n}_{11} a}{\sigma^2+p^{n}_{12} b})+log_2(1+\frac{ p^{n}_{12} c}{\sigma^2+p^{n}_{11} d})\\
\text{subject to}~~ &{{C}_{2}}:~p_{11}^{n}\ge 0 ~and ~p_{12}^{n}\ge 0\\
&{{C}_{3}}:~~~ \underset{n=1}{\overset{N}{\mathop \sum }}\,x_{ij}^{n}p_{ij}^{n}\le {{p}_{max}}\\
\end{aligned}
\end{equation}
Noting$log(1+A)log(1+B)$ is equivalent to $log(1+B+A+AB)$.thus, the optimization problem in (\ref{max_prob.5}) can be simplified to
\begin{equation}\label{max_prob.8}
\begin{aligned}
\max_{p^{n}_{11},p^{n}_{12}}&(\frac{ p^{n}_{11} a}{\sigma^2+p^{n}_{12} b})+(\frac{ p^{n}_{12} c}{\sigma^2+p^{n}_{11} d})+(\frac{ p^{n}_{11} a}{\sigma^2+p^{n}_{12} b})(\frac{ p^{n}_{12} c}{\sigma^2+p^{n}_{11} d})\\
\text{subject to}&~{{C}_{1}}:~p_{ij}^{n}\ge 0\\
&~~~{{C}_{2}}:\underset{n=1}{\overset{N}{\mathop \sum }}\,x_{ij}^{n}p_{ij}^{n}\le{{p}_{max}}\\
\end{aligned}
\end{equation}
Which can be reformulated as: :
\begin{equation}\label{max_prob.9}
\begin{aligned}
\max_{p^{n}_{11},p^{n}_{12}}&\Big(\frac{ (p^{n}_{11})^{2 }ad+(p^{n}_{11})^{2} a\sigma^2+(p^{n}_{12})^{2} bc+(p^{n}_{11})^{2} c\sigma^2+p^{n}_{11}p^{n}_{12} ac}{\sigma^4+p^{n}_{12} b\sigma^2+p^{n}_{11} d\sigma^2+p^{n}_{11}p^{n}_{12}bd}\Big)\\
 \text{subject to}~&{{C}_{1}}:~p_{ij}^{n}\ge 0\\
~~~&{{C}_{2}}:\underset{n=1}{\overset{N}{\mathop \sum }}\,x_{ij}^{n}p_{ij}^{n}\le{{p}_{max}}\\
\end{aligned}
\end{equation}
 To optimize the power of each user pair, we make use of the following lemma
\subsection{\textbf{Lemma1}}
 let's consider a nonlinear fractional programming problem:
 \begin{equation}\label{sysmod8}
   \max_{\textbf{x}}\frac{h(x)}{g(x)} ~\textbf{x}\in X \subseteq R^{n}
 \end{equation}
The objective function of (\ref{sysmod8}) is a fraction of two convex function called fractional problem. To address the optimal solution, we define the following function:
\begin{equation}\label{sysmod14}
  F(\mathbf{x};\lambda)=h(x)-\lambda g(x),
 ~\mathbf{x}\in X,~\lambda>0
\end{equation}
\begin{proof}
  We consider the following optimization problem.  
\begin{equation}\label{sysmod15}
    x(\lambda)={\arg max}_{x}F(x;\lambda)
  \end{equation}
  Also
 \begin{equation}\label{sysmod16}
    \pi(\lambda)={\max}_{x}F(x;\lambda)
  \end{equation}
  If there exist $\lambda*\geq 0$ for which $\pi(\lambda*)=0$,then $x*\equiv x (\lambda*)$ is an optimal solution of (\ref{sysmod8}).\\
  \textbf{Proof} : see \cite{b14}
\end{proof}
According to the Lemma1 and referring to (\ref{max_prob.9}), we define $F(\textbf{p};\lambda)$ as follows  :
\begin{eqnarray}\label{sysmod17}
    &F(\textbf{p};\lambda)=\{\max_{\textbf{p}}f(\textbf{p})-\lambda g(\textbf{p})~|\textbf{p}\in{P}\}\\
 \text{subject to}~& 0\leq P\leq p_{max}\nonumber
      \end{eqnarray}
   we define $f(\textbf{p}) $and $g(\textbf{p})$ where:
   \begin{equation}\label{sysmod18}
     f(\textbf{p})=\Big({ (p^{n}_{11})^{2 }ad+(p^{n}_{11})^{2} a\sigma^2+(p^{n}_{12})^{2} bc+(p^{n}_{11})^{2} c\sigma^2+p^{n}_{11}p^{n}_{12} ac}\Big)
   \end{equation}
    \begin{equation}\label{sysmod19}
     g(\textbf{p})=\Big(\sigma^4+p^{n}_{12} b\sigma^2+p^{n}_{11} d\sigma^2+p^{n}_{11}p^{n}_{12}bd\Big)
   \end{equation}
The problem in (\ref{max_prob.8}) is a standard fractional problem which can be solved by the Dinkelbach algorithm \cite{b16} in polynomial time. The algorithm summarized in proposition 1.\\
 \textbf{\emph{Proposition1:Optimality}}
 \begin{equation}\label{sysmod20}
   F({\lambda^{*}})=\{\max_{\textbf{p}}f(\textbf{p})-\lambda^{* }g(\textbf{p})~ |\textbf{p}\in{P}\}=0
 \end{equation}
 if and only if
 \begin{equation}\label{sysmod21}
 \lambda^{*}=\frac{f(\textbf{p}^{*})}{g(\textbf{p}^{*})}=\max\{\frac{f(\textbf{p})}{g(\textbf{p})}|\textbf{p}\in{P}\}
 \end{equation}
  \begin{algorithm}[t]
 \caption{Algorithm 1 Dinkelbach Algorithm}
1-Initialize $\lambda=0$ \\
2-set error tolerance $\delta\ll1$ and iteration index n=1\\
3-\textbf{Repeat}\\
4- $\textbf{p}^{*}=\{\max_{\textbf{p}}f(\textbf{p})-\lambda g(\textbf{p})\}$\\
5-$\lambda_{n+1}=\frac{f(\textbf{p}^{*})}{g(\textbf{p}^{*})}$\\
6-n=n+1\\
\textbf{7-Until }$f(\textbf{p})-\lambda g(\textbf{p})\leq \delta$ (convergence check)
  \end{algorithm}
 Based on Lemma1, the optimal power allocation can be obtained.\\
 To solve the step 4 of the this algorithm we will assume that the objecttive optimization problem of (11) can be define as follow:
  \begin{eqnarray}\label{sysmod21}
    \max_{\textbf{p}}f(\textbf{p})=\frac{\mathbf{p^TMp+c^Tp}+\alpha}{\mathbf{p^TQp+d^Tp}+\beta}\nonumber\\
    \text{subject to}~\textbf{p}\in{P}~\subseteq R^{n} , \textbf{p}\leq p_{max}
  \end{eqnarray}
  Where $M$ and $Q$ are $n*n$ symmetric positive semi-definite matrices.To address the optimal solution,we define the following function.
 \begin{eqnarray}\label{max_prob12}
   & F(\textbf{p};\lambda)=\max_{\mathbf{p}}\{{\mathbf{p^TMp+c^Tp}+\alpha}-\lambda_{n}(\mathbf{p^TQp+d^Tp}+\beta)|\mathbf{p}\in P\}\nonumber \\
   &=\textbf{p}^{T}(\textbf{M}-\lambda_{n}\textbf{Q})\textbf{p}+(c-\lambda_{n})^{T}\mathbf{p}+\alpha-\lambda_{n}\beta\\
   \text{subject to}~&\textbf{p}\in{P}~\subseteq R^{n} , \textbf{p}\leq p_{max}
  \end{eqnarray}
  To simplify we assume that $(\mathbf{M-\lambda_{n}Q})$are positive semi definite matrices. This hypothesis guaranties that the active set method for solving $F(\mathbf{p};\lambda)$  is finitely convergent.Dinkelbach algorithm is stated as follows:\\
 4-1 Let $\textbf{p}_{1}\in{P}$ be a fisible point of   and
 $\lambda_{1}=\frac{f(\textbf{p}^{*}_{1})}{g(\textbf{p}^{*}_{1})}$ Let  $n=1, s=1, \mathbf{H}_{n}=\mathbf{M-\lambda_{n}Q}$ , $c_{n}=c-\lambda_{n}d$ solve the following direction to find the problem as follows:\\
 4-2 $\mathbf{Q_{n}(p^{*})}=\min\{-\Big(\mathbf{y^{T}H_{n}y}+(c_{n}+H_{n}p^{*})\Big)\}:a^{T}_{i}\mathbf{y}=\mathbf{0}$
and denote by  $\mathbf{y_{s}}$ the optimum obtained\\
4-3 If  $\mathbf{y_{s}}=0$ let $v^{*}_{i}$ denote denote the optimum lagrange multipliers for $\mathbf{Q_{n}(p^{*})}$ if $v^{*}_{i}\geq 0$ then $p^{*}$ is optimal to $F(\textbf{p};\lambda)$ and go to convergence check\\
Else $\mathbf{y_{s}\neq 0}$, let\\
$\alpha_{s}=\frac{p_{max}-\mathbf{p^{*}}}{y_{s}}$\\
If $\alpha_{s}>1$ put $p^{*}_{s+1}=p^{*}_{s}+\mathbf{y_{s}}$ otherwise $p^{*}_{s+1}=p^{*}_{s}+\alpha_{s}\mathbf{y_{s}}$ \\
4-4 let s=s+1 and go to step 4-2
\section{Simulation result}\label{sysmulation result}
This section aims at comparing the performance of the proposed sub-channel assignment and power control mechanism in the uplink a two-cell network to that of proposed in the literature including what is reported in\cite{b3}. We suppose that the number of sub-channels as well as the number of users are three. The wireless channel has Rayleigh fading distribution.We assume that the channel gain are composed of the path loss and frequency selective Rayleigh flat fading according to the $h^{n}_{i(j)j}=\phi^{n}d^{-\alpha}_{i(j)j}$  where, d
is the distance between the $i^{th}$ user of the $j^{th}$ cell to the   $j^{th}$ Bs. The distance between each user and corresponding base station is 100m and the distance between it and another base station is 500m. and the normalized noise power is set to one.
We use Monte Carlo simulation in which the maximization problem is solved for many realizations of the channel and we represent the average results.Fig2 illustrates the average of sum rate versus SNR for different methods. Method A represents the optimal solution for which channel assignment is performed using exhaustive search and the optimal power control is done according to our proposed optimal power allocation policy. Method B depicts the proposed method for sub-channel assignment based on the Hungarian method and the proposed optimal power control. Method C indicates that sub-channel assignment and power control is done according to the iterative method addressed in~\cite{b3} through the use of DC-programming approach. the Hungarian method with full power assumption to the simulation results confirming our assertion regarding the advantage of the proposed method in mid SNR region. Moreover, the sub-channel assignment based on exhaustive search method when incorporating full power is also added to the figures, showing the advantage of power control mechanism for SNR values greater than zero.Finally last curve depict random selection strategy with full power control.
Although sub-channel assignment at the high SNR region is not optimal, the results indicate that it yields favorable result even at high SNR region. Moreover, the power allocation policy is optimal. As observed in Fig.2, at low SNR region, our proposed method outperforms the existing DC-programming and achieves higher data rates as we expected. Also, the average sum-rate of the proposed method coincides that of the exhaustive search with much lower complexity.

\begin{figure}
  \centering
  \includegraphics[width=16.00cm]{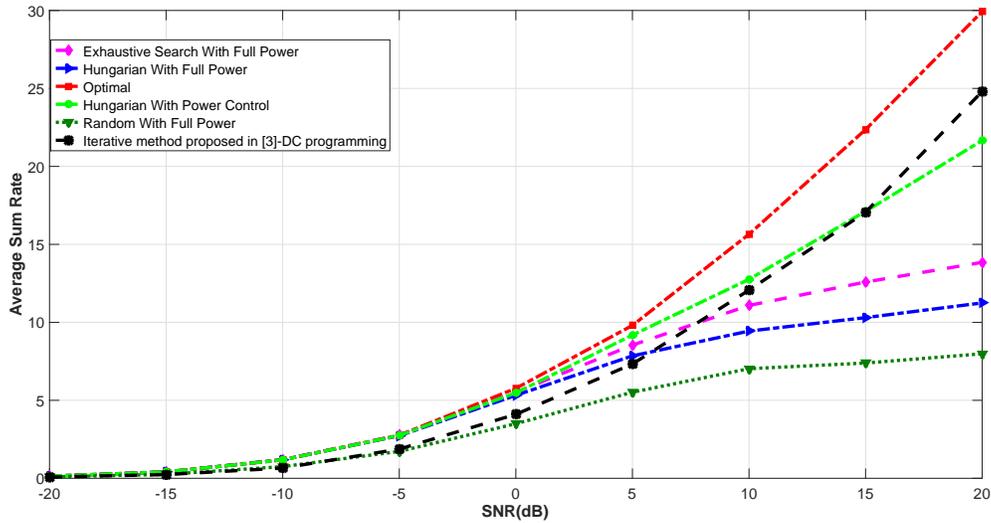}
  \caption{The average data rate of network for different method}
  \label{fig:ISC_Full}
\end{figure}
\section{Conclusion}\label{sec:conclusion}
This paper proposes a joint uplink sub-channel assignment and power control which is close-to-optimal at low SNR region. To this end, the optimization problem is divided in two steps. First, the sub-channel assignment is selected according to the Hungarian method and then the power of each user is devised through solving a fractional quadratic optimization problem. Numerical results indicate that at low to mid SNR region, the proposed method outperforms the existing methods addressed in the literature.

\end{document}